\documentclass[%
 reprint,
 superscriptaddress,
 amsmath,amssymb,
 aps,
 nofootinbib,
 prb,
 nolongbibliography,
]{revtex4-2}

\usepackage[utf8]{inputenc}
\usepackage[T1]{fontenc}
\usepackage{graphicx}
\usepackage{dcolumn}
\usepackage{bm}
\usepackage{hyperref}

\begin{document}

\preprint{APS}

\title{On the miscibility gap in tungsten-based alloys}

\author{Andrzej Piotr Kądzielawa}
\email{andrzej.piotr.kadzielawa@vsb.cz}
\affiliation{IT4Innovations, VŠB - Technical University, Ostrava-Poruba, Czechia}
\affiliation{Institute of Theoretical Physics, Jagiellonian University, Kraków, Poland}
\author{Dominik Legut}%
\email{dominik.legut@vsb.cz}
\affiliation{IT4Innovations \$ Nanotechnology Centre, CEET, VŠB - Technical University, 17.listopadu 2172/15, Ostrava, CZ 708 00 Czech Republic}

\date{\today}

\begin{abstract}
In this work we establish an approach to model miscibility gaps of alloys using statistical physics, lattice dynamics from first-principles calculations. We carefully calculate the entropy to include all processes introducing disorder to the system, i.e., combining the electronic, phononic, and configuration entropies. Furthermore we present our algorithm for generating Special Quasirandom Structures (SQS). We model the miscibility gap in tungsten - chromium and tungsten - molybdenum systems, obtaining the agreement with the experimental data. Furthermore, we propose an enhancement for the tungsten-chromium W\textsubscript{70}Cr\textsubscript{30} alloy with tantalum and hafnium, leading to the modified stabilization temperatures $T_S$, where the solid solution is miscible.
\end{abstract}

\maketitle

\section{Introduction}
\label{sec:introduction}
With the rapid search of carbon-neutral power producing facilities, the need for working fusion reactors is higher than ever before. Successful Ignition Events \cite{AbuShawareb2022,Zylstra2022}, followed by net-positive cycle \cite{Zylstra2022b}, increased the priority of research on the crucial elements of a fusion reactor, e.g., the so-called first walls. One family of compounds, namely tungsten-based alloys represent prospective candidates to replace pure tungsten, as the alloy's features (e.g., selfpassivation) make them resistant to the Loss of Coolant Accident (LOCA) with concomitant air ingress into the reactor vessel. Since tungsten undergoes volatile oxidation when in a temperature above $500^\circ$ C (the nuclear afterheat causes a surge of temperature up to $1200^\circ C$), threatening structural damage. In case of selfpassivating tungsten alloys, the oxidation of the additional element forms a robust oxide scale \cite{Litnovsky,GarciaRosales}. The problem being that the alloy operates as purposed while in extreme conditions (high temperature). On the other hand in normal conditions it is not thermodynamically stable, diffusing into W-rich and Cr-rich phases.

Finding this so-called \emph{miscibility gap}, the research community of 20\textsuperscript{th} century \cite{DenBroeder,DenBroeder2,Porter,Rietveld,Margaria,Greenaway,Greenaway2,McQuillan,Venkatraman,Naidu}, did not propose any large-scale applications for W-Cr binaries. In contrast, now, the \texttt{W-Cr} alloy is a candidate for the first wall of a fusion reactor vessel \cite{Tanure} (\emph{e.g.}, 620 m\textsuperscript{2} for the International Thermonuclear Experimental Reactor). The thermal stability of \texttt{W-Cr} alloys was already studied experimentally \cite{Vilemova,Vilemova2,Calvi,Sal}, showing  fast decomposition kinetics close to $1000^\circ$ C, \emph{i.e.}, near the LOCA conditions. In such a case, changes in the phase content and degradation of oxidation performance are inevitable.

The density-functional-theory-based \emph{ab initio} computational modelling  \cite{Hohenberg,Kohn,KohnNobel,Liechtenstein,Dudarev,Kotliar,Vollhardt} is now the state-of-the-art approach when high-throughput (hundreds of millions corehours) calculations are to be considered. DFT-based methods already have their input into designing novel materials \cite{Curtarolo,Zhang,Zhang2,Zhang3}. 

Treatment of alloys in real-space requires averaging over different randomly-generated structures. The so-called \emph{special quasi-random structures} (SQS) \cite{Zunger,Wei,Jiang} reduce the computational-complexity by providing us with a set of most \emph{"alloy-like"} supercells.

In effort to decrease the closing temperature for the miscibility gap (henceforth referred to as a stabilization temperature $T_S$), we consider the enhanced W-Cr-X alloy. For the purpose of this paper we chose two refractory metals, \emph{i.e.}, hafnium (Hf) and tantalum (Ta). In Section~\ref{sec:model} we discuss the thermodynamics behind our calculations (Sec.~\ref{ssec:thermodynamics}), then the problem of geometry of the thermodynamical potential in composition diagram (Sec.~\ref{sssec:EoF}). Next we briefly focus on the generation of the SQSs (Sec.~\ref{ssec:sqs}) and other computational details (Sec.~\ref{ssec:compdeets}). In Section~\ref{sec:results}, we discuss the presence of the miscibility gap, it closure in high temperature (Sec.~\ref{ssec:miscibility}), and the impact on the $T_S$ of a small ($\sim$ 2 atomic \%) enhancement of an W\textsubscript{70}Cr\textsubscript{30} alloy with Hf and Ta(Ssec.~\ref{ssec:enhancement}).

\section{Model}
\label{sec:model}
We tackle the problem of thermodynamics of a solid solution, by starting from the choice of the Canonic Ensemble, with the Gibbs Free energy as our potential.
\subsection{Gibbs energy}
\label{ssec:thermodynamics}
\begin{figure}
    \centering
    \includegraphics[width=\linewidth]{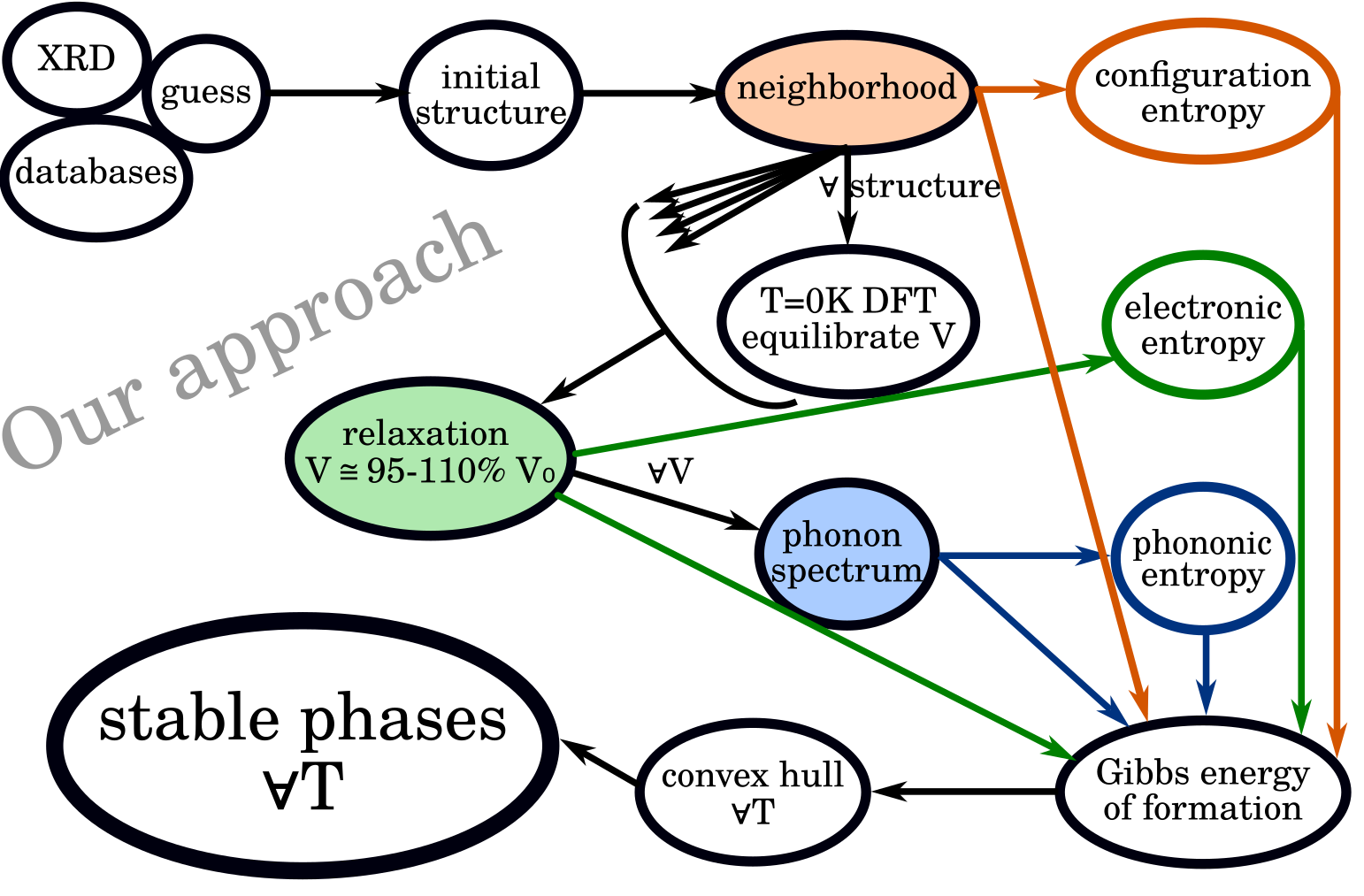}
    \caption{A general scheme for the computational approach. Note that the input is not required to be specific.}
    \label{fig:plan}
\end{figure}

We assume that \emph{(i)} we can model the electronic structure of an alloy in $T \neq 0 \, K$ by artificially expand the volume; \emph{(ii)} we obtain the temperature dependence of volume by minimizing the Helmholtz free energy; \emph{(iii)} the entropy is a sum of configuration, electronic, and phononic entropies, i.e.,
\begin{align}
    S\big(T\big) = S_\text{configuration} + S_\text{electronic}(T) + S_\text{phononic}(T),
\end{align}
note that the volume is a function of temperature (not an independent variable), as our thermodynamic function is the Gibbs energy $G(T,p)$ (here in $p=0$), not the Helmholtz Free energy $F(T,V)$. \emph{Cf.} the graphical scheme in Fig.~\ref{fig:plan}, where the steps required by configuration entropy are in orange, electronic entropy in green, and phononic entropy in blue.

An exemplary alloy $A_{x_1}B_{x_2}C_{x_3}\dots$ ($\sum_i x_i = 1$) has the configuration entropy
\begin{align}
S_\text{configuration} \equiv - k_{B} \sum_{i} x_i \log (x_i).
\end{align}
For more details on various entropy calculations see Appendix~\ref{app:gibbs}.

\subsubsection{Gibbs energy of formation}
\label{sssec:EoF}

To analyze the instability - the likelihood to diffuse into two (or more) subsystems, we start from the Gibbs Energy of Formation (for an exemplary compound $A_{x}B_{y}C_{z}$)
\begin{align}
\Delta G_{A_{x}B_{y}C_{z}} \equiv  G_{A_{x}B_{y}C_{z}} - xG_A - yG_B - zG_C,\\
\text{ where } x+y+z \equiv 1,
\end{align}
while $G_A\equiv G_A(T,p)$, $G_B\equiv G_B(T,p)$, and $G_C\equiv G_C(T,p)$ are the Gibbs free energies of pure A, B, and C systems respectively. If we consider a vector space: the composition ($x,y$, since $z\equiv z(x,y)$)supplemented with the energy of formation ($\Delta G$), the point in that space represents a single real-space system. We identify a stable point when its $\Delta G$ is negative and lower than a that of a linear combination of its neighbors energies of formations $\big\{\Delta G\big\}_{N}$ (such as the linear composition is equated to the point composition). Thus, the problem of stability of a set of compounds on the composition-energy of formation diagram reduces to a problem of finding the convex hull of the set of points representing the physical systems.

\subsection{Special Quasirandom Structures}
\label{ssec:sqs}
Proposed in the early '90 \cite{Zunger}, the special quasirandom structures (SQS) are now the established technique to model dynamics of an alloy. They stem from the principle that the best crystalic structure to represent an alloy A\textsubscript{x}B\textsubscript{y}C\textsubscript{z}\dots will have the probability of finding an atom A on any coordination sphere to be exactly x (etc. for B - y,\dots). Normally, generation of an SQS starts from some random selection and then via the Metropolis-Hastings algorithm \cite{Metropolis,Hastings} optimizes the error function. This approach, while time and memory conserving, produces structures from 1.P1 symmetry group, since all the other structures consists a null set in the space of all possible configurations.

Our approach to determine a fast SQS is to \emph{(i)} create a starting set from the single-element lattice with $N$ nodes (body-centered cubic - bcc - in this case with tungsten), and \emph{(ii)} randomly change $n \ll N$ atoms on the starting element lattice, \emph{(iii)} analyze symmetry of the resultant set to remove redundant structures, as well as the ones with 1.P1 symmetry, and finally \emph{(iv)} calculate error function for each surviving structure. The remaining set constitutes a population. Each random structure is associated with a number (the error function). If the concentration is the one we are looking for, we have a ranked list of structures and we quit this algorithm. If not, we employ \emph{(v)} extinction, \emph{i.e.}, i\textsuperscript{th} structure has a probability to survive
\begin{align}
    P_i \equiv \frac{\mathfrak{E}_0}{\mathfrak{E}_i},
\end{align}
where $\mathfrak{E}_0$ is the lowest value of error function (\emph{i.e.,} $\min\big(\big\{\mathfrak{E}_i\big\}\big)$). Next, we \emph{(vi)} create a new set of $M$ best surviving structures and use it as an input in point \emph{(ii)}, until expected concentration is achieved.
\begin{figure*}
    \centering
    \includegraphics[width=\linewidth]{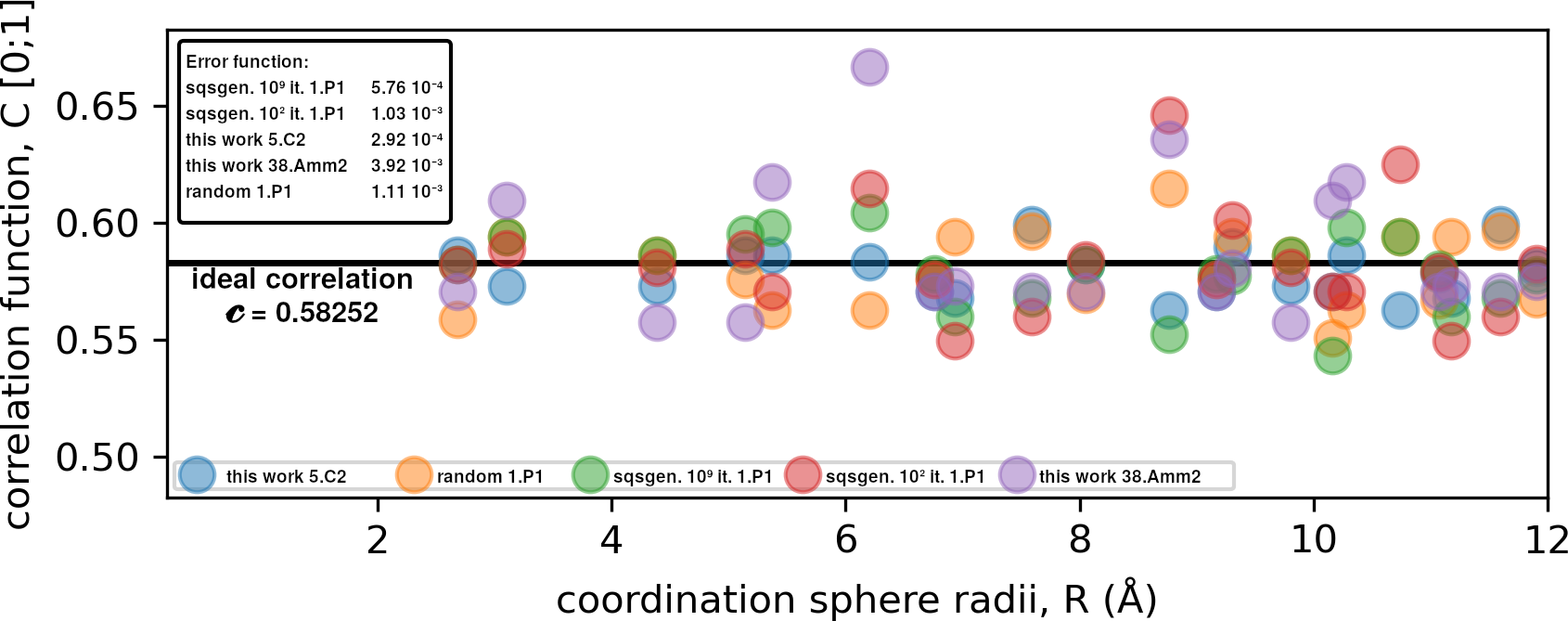}
    \caption{Correlation function for different Special Quasirandom W\textsubscript{70}Cr\textsubscript{30} Structures. We used two \texttt{sqsgenerator} \cite{sqsgen} generated 1.P1-symmetry cells with 10\textsuperscript{9} and 10\textsuperscript{2} iterations. We included our own cells from 5.C2 and 38.Amm2 symmetry groups, while the result for purely random 1.P1 cell is given as a reference point. Inset contains calculated values of the error function $\mathfrak{E}$.}
    \label{fig:correlation}
\end{figure*}

This approach provides comparable results to the established tool, namely \texttt{sqsgenerator} \cite{sqsgen}, correlation-wise (see Fig.~\ref{fig:correlation}). \emph{E.g.}, our predictions of high-temperature magnitudes of elastic moduli are close to the experimental results (for both see \cite{Veverka}).

\subsection{Computational details}
\label{ssec:compdeets}
Using the Vienna ab initio simulation package (VASP) \cite{VASP} we performed the electronic calculations to equilibriate volume in $T=0K$. For that we used PAW \cite{PAW} pseudopotentials supplied with VASP and the Perdew-Burke-Ernzerhof (PBE) generalized gradient approximation (GGA) exchange-correlation functional \cite{PBE}. To model the temperature dependence on the crystals representing W-Cr alloy, we employ the quasiharmonic approximation \cite{QHA}, i.e., we compute the phononic spectra for volumes 90-110\% of the $T=0K$ volume $V_0$. We use PHONOPY \cite{PHONOPY} to find the minimal number of atomic displacements and analyze the output files.

For VASP calculations we used pseudopotentials with 6 valence electrons for W, Mo, and Cr, 5 for Ta, and 4 for Hf. We chose $3 \times 3 \times 3$ BCC supercell (i.e., with 54 lattice nodes). The k-point mesh was set to $4 \times 4 \times 4$  $\Gamma$-centered, where preliminary calculations showed energy saturation. The kinetic energy cut-off for the electrons was set to $320 eV$. We carried out the frozen-ion calculations with a 0.1 Å displacement, and we used a $69 \times 69 \times 69$ reciprocal cell mesh for estimation of the thermal properties up to 3000 K with a 1 K step.

\section{Results}
\label{sec:results}
As aforementioned, in our approach there is no ab-initio adjustable parameters. Below, we discuss the results obtained using our systematical approach.

\subsection{Miscibility gap}
\label{ssec:miscibility}
\begin{figure}[t]
    \centering
    \includegraphics[width=\linewidth]{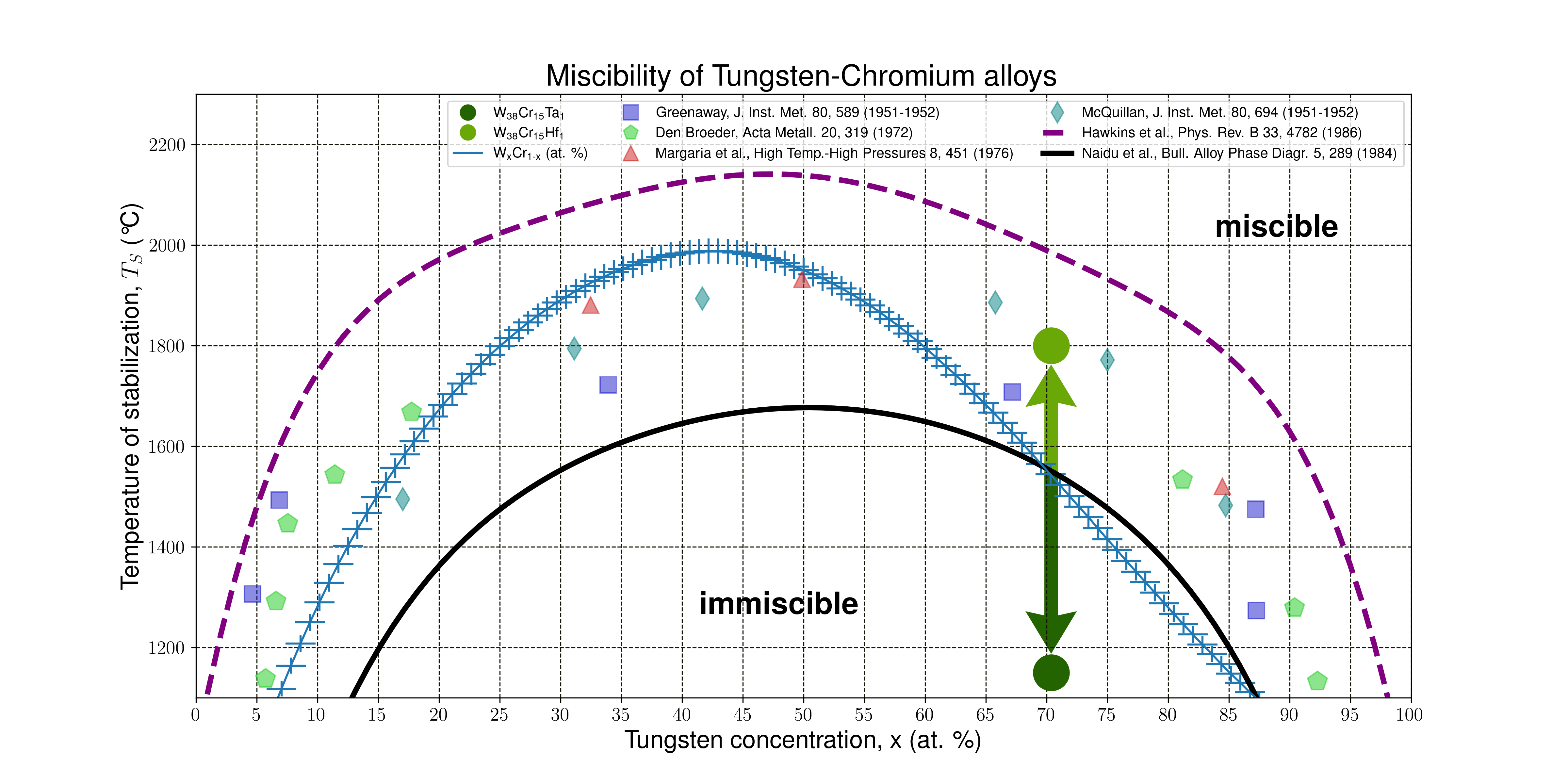}
    \caption{Tungsten - chromium miscibility, doping with Tantalum (Green) and Hafnium (Lime) modifies stabilization temperature.}
    \label{fig:WCrT}
\end{figure}

\begin{figure}
    \centering
    \includegraphics[width=\linewidth]{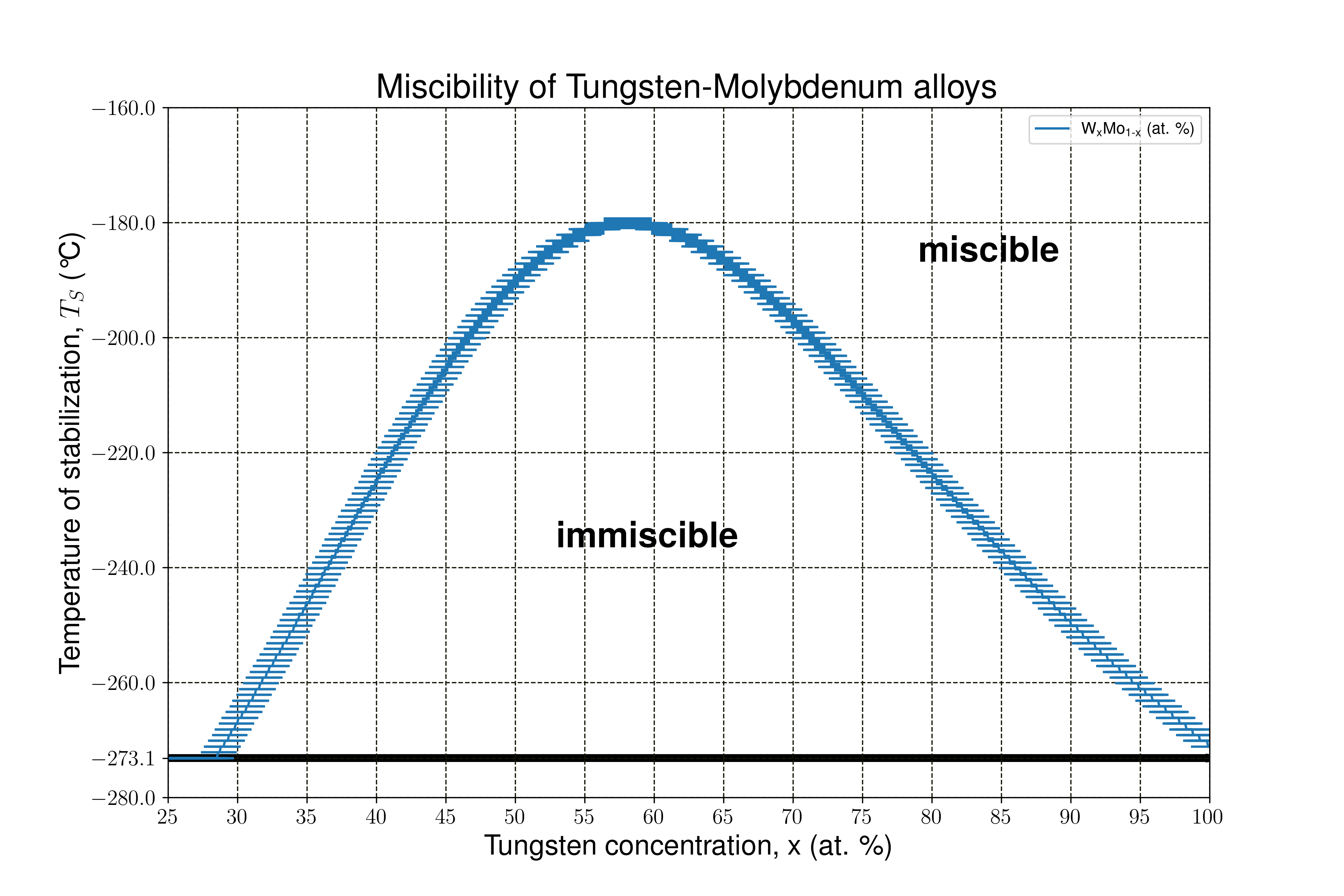}
    \caption{Tungsten - Molybdenum miscibility. Solid line marks the absolute zero ($0 \ K \equiv -273.15 \ ^\circ C$.}
    \label{fig:MoWT}
\end{figure}

To assess the prediction capabilities of our method, we reproduce the high-temperature miscibility gap of W-Cr binary alloys (vide Fig.~\ref{fig:WCrT}) and a low-temperature miscibility gap of W-Mo binary alloys (vide Fig.~\ref{fig:MoWT}).
Our results (blue crosses in Fig.~\ref{fig:WCrT}) lay closer to the experimental points when compared to the previous parameter-adjustable results \cite{Hawkins,Naidu} (note that the variance of the experimental points is of the same magnitude as our results error).

To be sure that a large miscibility gap in tungsten-chromium system is not a coincidence we also performed the same calculations for the tungsten-molybdenum system. The W-Mo system behave in the same \emph{qualitative} manner as W-Cr - there is a dome-shaped misciblity gap with a maximum stabilization temperature ($T_S$) for an atomic concentration close to 50-50 (atomic \%). There is yet a quantitative alteration - the temperature below which the solid solution becomes immiscible is expected to be two orders of magnitude smaller - a result we unequivocally obtain (\emph{cf.}, Fig.~\ref{fig:MoWT}).

\subsection{Enhancing W-Cr alloy}
\label{ssec:enhancement}
As previously stated, we are interested in decreasing of the stabilization temperature $T_S$ for a W\textsubscript{70}Cr\textsubscript{30} (atomic \%) alloy, by the means of enriching the system with small amount ($\sim$2 atomic \%) of other (refractory) metals. Here we consider hafnium and tantalum as such candidates. We use the same approach as described in Sec.~\ref{ssec:thermodynamics} to reliably study a convexity of a point (corresponding to the composition in question) in the composition -- energy of formation diagram, we generate (as we did in Sec.~\ref{ssec:sqs}) a set of compositions neighboring our point -(W\textsubscript{70}Cr\textsubscript{30})\textsubscript{98}X\textsubscript{2} - up to the third nearest neighbor. The analysis of the geometry in three-dimensional composition -- energy of formation space, shows decisively different behavior - increase of $T_S:\,1550 \,^\circ C \rightarrow\, 1800\,^\circ C$ for 2 \% Hf and decrease of $T_S:\,1550 \,^\circ C \rightarrow\, 1150\,^\circ C$ for 2 \% Ta, a result consistent with previous experimental measurements \cite{Veverka2}.

\section{Conclusion}
\label{sec:conclusion}
In this paper we have presented our approach to decrease the computational load for complicated problem like temperature dependence of alloy stability. Since we have incorporated our method of generating special quasirandom structures with some residual symmetry left\footnote{A note is here to make. No real-space supercell structure is perfectly disordered - translational symmetry always prevails. So from that point of view, considering systems with additional - though still low - symmetry does not introduce \textbf{new qualitative} error.} we are able to calculate whole phase diagram, close to experiment (vide Fig.~\ref{fig:WCrT}) or even modify parts of this diagram by enhancing alloy with Hf (increasing the miscibility gap) and Ta (decreasing the miscibility gap). These results are in agreement with previous experimental work \cite{Veverka2}. Our approach is parameterless and general and can be applied to other problems as well.

\section{Acknowledgements}
\label{sec:acknowledgements}
Financial support by the Czech Science Foundation through grant No. 20-18392S is acknowledged as well as the project e-INFRA CZ (ID:90140) by Czech MŠMT.

\newpage

\nocite{*}
\bibliography{bibliography}

\newpage
\onecolumngrid
\appendix
\section{Gibbs energy}
\label{app:gibbs}
\begin{align}
    S = S_\text{configuration}+ S_\text{electronic}+ S_\text{phononic}.
\end{align}
With $\beta \equiv \tfrac{1}{k_B T}$, an exemplary alloy $A_{x_1}B_{x_2}C_{x_3}\dots$ ($\sum_i x_i = 1$) has the configuration entropy
\begin{align}
S_\text{configuration} \equiv - k_{B} \sum_{i} x_i \log (x_i),
\end{align}
while the electronic entropy
\begin{align}
 S_\text{electronic} \equiv - k_{B} \sum_{\sigma} \int d\epsilon DOS_{\sigma} (\epsilon)\bigg( f_\text{FD}(\epsilon) \log f_\text{FD}(\epsilon) + [1-f_\text{FD}(\epsilon)] \log [1-f_\text{FD}(\epsilon)] \bigg),
\end{align}
where $\sigma$ corresponds to spin, and $ f_\text{FD}(E) \equiv \left(  \exp [ (E-E_\text{Fermi})\beta ] +1 \right) ^{-1} $ is the Fermi-Dirac distribution. Next, the phononic term
\begin{align}
S_\text{phononic} \equiv& - k_B \sum_{\mathbf{q},\nu} \bigg( \log (f_\text{BE}(\hbar \omega_{\mathbf{q},\nu})) - \hbar \omega_{\mathbf{q},\nu} \beta f_\text{BE}(\hbar \omega_{\mathbf{q},\nu}) \exp (- \hbar \omega_{\mathbf{q},\nu} \beta) \bigg)\,
\end{align}
where $\omega_{\mathbf{q},\nu}$ is the angular frequency of the $\nu$-phonon at $\mathbf{q}$ direction in the reciprocal space. $f_\text{BE}(E) \equiv \left(  \exp ( E\beta ) -1 \right) ^{-1}$ is the Bose-Einstein distribution.

\end{document}